# Fast electronic relaxation in metal nanoclusters via excitation of coherent shape deformations: Circumventing a bottleneck

Vitaly V. Kresin[1] and Yu. N. Ovchinnikov[2]

[1] *Department of Physics and Astronomy, University of Southern California, Los Angeles, California 90089-0484*

[2] *L. D. Landau Institute for Theoretical Physics, Russian Academy of Sciences, 119334 Moscow, Russia*

## ABSTRACT

Electron-phonon relaxation in size-quantized systems may become inhibited when the spacing of discrete electron energy levels exceeds the magnitude of the phonon frequency. We show, however, that nanoclusters can support a fast nonradiative relaxation channel which derives from their distinctive ability to undergo Jahn-Teller shape deformations. Such a deformation represents a collective and coherent vibrational excitation and enables electronic transitions to occur without a multiphonon bottleneck. We analyze this mechanism for a metal cluster within the analytical framework of a three-dimensional potential well undergoing a spheroidal distortion. An expression for the time evolution of the distortion parameter is derived, the electronic level crossing condition formulated, and the probability of electronic transition at a level crossing is evaluated. An application to electron-hole recombination in a closed-shell aluminum cluster with 40 electrons shows that the short (~250 fs) excitation lifetime observed in recent pump-probe experiments can be explained by the proposed mechanism.

# I. INTRODUCTION

One of the central challenges in the physics of clusters and related nanoscale systems is the issue of the relaxation dynamics of excited states. In particular, what are the specific decay channels and rates of single-particle and collective electronic excitations in size-quantized structures, and how do they evolve from the molecular limit of vibronic coupling to the electron-phonon interaction characteristic of the bulk?

The so-called "phonon bottleneck" problem[1] is noteworthy. Consider a small particle with electrons occupying a set of discrete energy levels, up to some highest occupied level $|A\rangle$ (analogous to the "highest occupied molecular orbital," or "HOMO," in spectroscopic language). Suppose an electron is excited into the next higher level, $|B\rangle$ (analogous to the "lowest unoccupied molecular orbital," or "LUMO"); can it now undergo nonradiative relaxation? In a conventional solid-state system, electron-hole recombination can be efficiently accomplished by phonon emission. But in a cluster the gap $E_B$-$E_A$ (the "intershell spacing," or the "HOMO-LUMO gap") can easily exceed the scale of vibrational energies by a very large factor. Thus to bridge the gap an electron would need to emit a multitude of vibrational quanta simultaneously, a high-order process of exceedingly low probability.

In this paper, we point out and analyze an efficient relaxation mechanism which is based on the fact that free nanoclusters possess an important degree of freedom: they can undergo significant shape deformations. This feature distinguishes them from constrained nanostructures such as semiconductor quantum dots. The proposed mechanism is illustrated in Fig. 1: upon excitation, the cluster sets out on a Jahn-Teller distortion from its original spherical shape; the energies of the $A$ and $B$ orbitals shift and eventually cross; an intershell transition occurs; and finally the electron lands back in its original shell and the shape oscillation decays into a swarm of incoherent vibrations (heat). An essential point to note is that this process is not subject to the phonon bottleneck issue: shape deformation is a coherent state, i.e., a coherent multiphonon excitation without any additional smallness. The process is analogous to internal conversion at an avoided crossing in polyatomic molecules,[2] with the particularity that it involves a distinctly collective coordinate.

The general treatment will be supplemented by a specific illustration referring to a recent time-resolved two-photon photoemission experiment on free $Al_n^-$ clusters in a beam.[3,4] Aluminum clusters exhibit electronic shell structure,[5] and $Al_{13}^-$ is a "magic" cluster: its 40 valence electrons are accommodated in closed shells (*1s, 1p, 1d, 2s, 1f, 2p*), and a substantial gap separates the highest occupied level (*2p*, corresponding to the label *A* above), from the next, lowest unoccupied one (*1g*, corresponding to *B*). In the experiment, a femtosecond laser pulse resonantly excited an electron from *A* into *B*,[6] and a subsequent ionizing pulse probed the population of the excited level after a certain delay. A surprising observation was that the magic $Al_{13}^-$ cluster had a relaxation rate as fast as neighboring non-magic clusters (estimated at about 250 fs), despite its considerably larger excitation gap. This implies that electron-electron scattering is not the central factor, and indicates "the existence of a very effective relaxation mechanism, which is independent on the electronic structure."[3]

So if there are no available intermediate electronic states in the gap, and radiative decay is known to occur on much longer time scales, a natural deduction is that the electron must relax via strong electron-phonon coupling. But this evokes the aforementioned phonon bottleneck issue: the excitation gap in $Al_{13}^-$ is 1.5 eV,[3] while the phonon energy in Al is much smaller: ~40 meV.[8]   We will demonstrate that the coherent deformation mechanism can explain the experimentally observed time scale.

Below, we treat the process step-by-step via an analytical model calculation for a finite square-well potential box filled with electrons, one of which is in an excited state..  Section II calculates the deformation-induced shift and crossing of the uppermost electronic levels, Section III evaluates the time needed to reach the level-crossing point, and Section IV considers the transition probability at this point.   Quadrupole and octupole deformations are invoked and analyzed.

## II. LEVEL SHIFTS AND CROSSINGS UNDER THE INFLUENCE OF QUADRUPOLE SHAPE DEFORMATIONS

As stated above, we model the cluster electrons as a Fermi gas confined to a spherically symmetric square-well potential.  This is, of course, only an approximation to more accurate self-consistent shell-model potentials, but the qualitative character of the low-lying shells in clusters and nuclei is not very sensitive to the well shape.[9]  The wave functions and energy levels of electrons in such a potential are given by

$$\psi_{nlm} = c j_l\left(k_{nl}r\right) Y_{lm}(\theta, \phi); \quad E_{nl} = \hbar^2 k_{nl}^2 / (2m^*) . \tag{1}$$

Here $c$ is the normalization constant, $j_l$ are spherical Bessel functions, $Y_{lm}$ are spherical harmonics, $k_{nl}$ is the $n^{th}$ root of the equation $j_l(k_{nl}R)=0$, $R=r_s a_0 N_e^{1/3}$ is the cluster radius ($r_s$ is the Wigner-Seitz parameter, $a_0$ is the Bohr radius, $N_e$ is the number of valence electrons in the cluster), and $m^*$ is the electron effective mass.[10]

Now suppose the particle surface proceeds to distort in an axially symmetric manner parametrized by a set of deformation parameters[13,14] $\alpha_L$:

$$R' = R\left[1 + \sum_L \alpha_L P_L(\cos\theta)\right] \equiv R\left[1 + \sum_L f_L(\theta)\right]. \qquad (2)$$

This, of course, shifts the electronic energy levels. To calculate the shift for a deformation of some particular multipolarity $L$ we introduce a rescaled radial coordinate $\tilde{r} = r[1 + f_L(\theta)]^{-1}$. In terms of the spherical coordinates $(\tilde{r}, \theta, \phi)$, the boundary remains a sphere of radius $R$, but in the Hamiltonian there appears a correction to the Laplace operator: $\hat{\tilde{H}} = \hat{\tilde{H}}_0 + \hat{\tilde{H}}'_L$. Here $\hat{\tilde{H}}_0 = -\hbar^2\tilde{\nabla}^2/(2m^*)$, the tilde denotes the fact that the Hamiltonian and the wave functions will now be written in the "squeezed" coordinate system. To first order in the deformation, the perturbation is

$$\hat{\tilde{H}}'_L = -2f_L\hat{\tilde{H}}_0 + \frac{\hbar^2}{2m^*}\frac{1}{\tilde{r}}\frac{\partial}{\partial\tilde{r}}\left[\frac{\partial}{\partial\cos\theta}\left(\sin^2\theta\frac{\partial f_L}{\partial\cos\theta}\right) + \sin^2\theta\frac{\partial f_L}{\partial\cos\theta}\frac{\partial}{\partial\cos\theta}\right]. \qquad (3)$$

As a consequence, the shell degeneracy is removed and the energy levels split as follows:

$$E'_{nlm} = E_{nl} + \delta E_{nlm}, \qquad (4)$$

where

$$\delta E_{nlm} = \left\langle nlm\left|\hat{\tilde{H}}'_L\right|nlm\right\rangle = -2\alpha_L E_{nl}\left\langle nlm\left|P_L(\cos\theta)\right|nlm\right\rangle. \qquad (5)$$

The right-hand side of Eq. (5) arises from the fact that only the first term in Eq. (3) contributes to the diagonal matrix element. For quadrupole deformations ($L$=2) a calculation of Eq. (5) leads to the known result[15,16]

$$\delta E_{nlm} = \alpha_2 E_{nl} \eta(l, m), \tag{6}$$

where

$$\eta(l, m) = 2 \frac{3m^2 - l(l+1)}{(2l-1)(2l+3)}. \tag{7}$$

This specific expression has been derived for the square potential well model, but it will be qualitatively applicable to other shapes of the confining potential as well. For example, for a harmonic oscillator potential the shell energy shift differs only by a factor of two.[15]

From Eqs. (6),(7) it follows that to first order in the deformation parameter, the total energy of a filled shell doesn't change: $\sum_{-l \leq m \leq l} \delta E_{nlm} = 0$. This reflects the fact that for a closed-shell cluster the spherical shape represents a minimum-energy configuration (at least a local minimum). It is the presence of incompletely filled shell levels that drives cluster shape deformations.

This is the case in our situation: an electron promoted from the originally filled $A$ shell into the originally empty $B$ shell. The cluster will begin to deform until the $A$ sublevel containing the hole (call it $|n_A, l_A, m\rangle$) approaches the $B$ sublevel ($|n_B, l_B, m\rangle$) containing the excited electron, at which point an interlevel transition can occur. (Transitions will take place only between levels with the same value of $m$, hence both wave functions contain the same index.) In other words, the critical deformation parameter $\alpha_{2,cr}$ for level crossing is determined by setting

$$\delta E_{el} \equiv \delta E_{n_B, l_B, m} - \delta E_{n_A, l_A, m} \tag{8}$$

equal to the shell gap ($E_B$-$E_A$). The negative sign of the second term arises because the total energy of the remaining occupied $A$ levels decreases by the same amount by which the hole energy increases (since the total energy of a filled shell must remain unchanged). From Eq. (6), the result is

$$\alpha_{2,cr} = \frac{E_B - E_A}{\eta(l_A,m)E_A - \eta(l_B,m)E_B} .$$

(9)

Consider the $Al_{13}^-$ cluster example. For Al, $r_s$=2.1, $m^* \approx 1.4 m_e$.[8] For a cluster of 40 electrons, $R \approx 3.7$Å. The relevant roots of $j_l$ are $(k_{n=2,l=1}R)$=7.73 and $(k_{n=1,l=4}R)$=8.18, which translates into $E_{1g} \approx 1.12 E_{2p}$, $E_{2p} \approx 11$ eV, $E_{1g}$-$E_{2p} \approx 1.3$ eV. The latter value is in sensible agreement with the experimental gap magnitude of 1.5 eV.[3] The specific sublevels involved in the relaxation process can be identified from Eq. (6). Quadrupole distortion will split the *2p* level into two groups: $m$=0 will shift downwards, and $m=\pm1$ will shift upwards towards the *1g* shell according to $\delta E_{2p,m=\pm1} = \frac{2}{5}\alpha_2 E_{2p}$. The hole will "float up" this branch towards the photoexcited electron, which in turn will be "sliding down" along the $\delta E_{1g,m=\pm1} = -\frac{34}{77}\alpha_2 E_{1g}$ branch of the *2g* shell.[17] Put another way, the net change in the electronic energy, Eq. (8), will be $\delta E_{el} = -2\alpha_2 \left( \frac{1}{5} E_{2p} + \frac{17}{77} E_{1g} \right)$. Using the above relation between $E_{2p}$ and $E_{1g}$, we find that the relevant level crossing will occur at $\alpha_{2,cr} \approx 0.15$. This value of the deformation parameter agrees to within $\approx 10\%$ with that found from the Clemenger-Nilsson diagram of electronic levels in spheroidal metal clusters.[18] The diagram also illustrates that the linear approximation for $\delta E_{nlm}$ holds well for many subshells up to rather high values of the distortion parameter.

Now that we have found the point at which the electron and hole curves cross and recombination can occur, two more questions must be answered: (1) how long after the electron excitation event (for our purposes, instantaneous) will the deformation coordinate reach this value, and (2) what is the transition probability at the crossing point? These questions are taken up in the following two sections.

### III. CLUSTER SHAPE OSCILLATIONS

The deformation dynamics of the confining potential well, which models the massive ionic core, may approximately be treated classically. To determine the low energy cluster shape oscillation spectrum, we therefore need to include a term describing the potential energy of volume-conserving surface deformations of an elastic spherical crystallite (for example, the clusters analyzed in the experiment [3] are expected to be below their melting point). For cubic crystals, the elastic energy density is[8,19]

$$U = \tfrac{1}{2} C_{11}(u_{xx}^2 + u_{yy}^2 + u_{zz}^2) + C_{12}(u_{xx}u_{yy} + u_{xx}u_{zz} + u_{yy}u_{zz}) + 2C_{44}(u_{xy}^2 + u_{xz}^2 + u_{yz}^2), \quad (10)$$

where $u$ are components of the strain tensor, and $C$ are the elastic moduli.

For quadrupolar shape distortions, one finds (see Appendix A) that the elastic potential energy is determined only by the following combination:

$$E_{pot} = \pi R^3 \alpha_2^2 (C_{11} - C_{12}). \quad (11)$$

Here $\alpha_2$ is the shape deformation parameter introduced in the previous section. (In principle, deformation of a cluster ion also gives rise to Coulomb potential energy, but in the present case the Coulomb energy[20] is negligible compared with the elastic energy.)

Finally, the kinetic energy of the quadrupole surface oscillation is given by[13]

$$E_{kin} = (\pi / 5) \rho R^5 \dot{\alpha}_2^2, \quad (12)$$

where $\rho$ is the density and $\dot{\alpha}_2 \equiv \partial \alpha_2 / \partial t$.

We can now write down a general equation expressing energy conservation for a cluster undergoing small-amplitude spheroidal shape deformations: $E_{kin} + E_{pot} + \delta E_{el} = 0$, or

$$\left( \pi / 5 \right) \rho R^5 \dot{\alpha}_2^2 + \pi R^3 \alpha_2^2 (C_{11} - C_{12}) + \alpha_2 \left[ \partial E_{el} / \partial \alpha_2 \right]_0 = 0 \,. \qquad (13)$$

The first two terms are the kinetic and potential energies of deformation, Eqs. (11) and (12), and the third terms is the concomitant change in the electronic energy (the derivative is evaluated at $\alpha_2 = 0$). The initial conditions for our situation are $\alpha_2(t=0)=0$, $\dot{\alpha}(t=0)=0$: at the instant of electronic excitation, the cluster core has not yet started moving away from its original spherical shape. This differential equation has the solution

$$\alpha_2(t) = a \sin^2 \left( \tfrac{1}{2} \Omega t \right), \qquad (14)$$

where the characteristic shape oscillation frequency is

$$\Omega^2 = 5 \frac{C_{11} - C_{12}}{\rho R^2} \qquad (15)$$

and the oscillation amplitude is

$$a = \frac{-\left[ \partial E_{el} / \partial \alpha_2 \right]_0}{\pi R^3 (C_{11} - C_{12})} \,. \qquad (16)$$

Eq. (14) is one of the main results: it describes the manner and the time scale of Jahn-Teller deformation of cluster shapes. It applies to liquid-drop clusters as well as to crystalline ones: in the former case the elastic energy term in Eq. (13) is replaced by a surface tension term, but the $\alpha_2$ dependence remains the same.[21]

In the specific case of a single electron-hole pair excitation Eq. (8) applies, and from Eq. (6) we obtain for the numerator of Eq. (16):

$$-\left[\partial E_{el}/\partial \alpha_2\right]_0 = E_{n_A,l_A,n}\eta(l_A,m) - E_{n_B,l_B,m}\eta(l_B,m) \ . \tag{17}$$

Considering again $\overline{\text{Al}_{13}}$, we substitute the parameters from the end of Sec. II together with the aluminum density and bulk moduli,[8] and find for this cluster: $\Omega \approx 3 \times 10^{13}$ s$^{-1}$ and $a \approx 0.2$. Using Eq. (14), this means that the first approach to the level crossing point,

$$\alpha_2(\tau) = \alpha_{2,cr} \tag{18}$$

will occur in $\tau \approx 100$ fs.

It remains to verify that the probability of an electronic transition at this point is not too small.

## IV. TRANSITION PROBABILITY AT THE CROSSING POINT

The picture so far is as follows: after an electron is transferred into the lowest unoccupied orbital, the cluster begins to undergo a quadrupole deformation according to Eq.(14), and the electron and hole energy levels approach each other at the point $\alpha_{2,\text{cr}}$. Here the excited electron can return into its original shell. The crossing terms are also commonly referred to as the "diabatic potential curves."[22] As the crossing point is passed at a certain speed $v$, the transition probability $w$ for a single passage can be evaluated by the Landau-Zener formula[23]

$w = \left\{1 - \exp\left[-2\pi V^2 / \left(\hbar v |F_A - F_B|\right)\right]\right\}$. Here $V$ is the coupling matrix elements of the two electronic wave functions at the crossing point, and $F$ are the forces (i.e., the slopes of the two crossing curves, $A$ and $B$ in our notation) at the same point.

For some cluster sizes, the sublevels of interest are directly coupled by the spheroidal deformation operator $\hat{\tilde{H}}_2'$ [Eq. (3)], in which case the above expression for $w$ can be applied immediately, with $V$ of the form $\langle A|\hat{\tilde{H}}_2'|B\rangle$. The exponent is likely to be rather large, and the transition probability near unity. This means that it will be possible to associate the electronic relaxation time with the time needed to reach the crossing point, i.e., with the root of Eq. (18).

However, there will commonly arise situations when direct coupling is absent. For example, in the example of photoexcited $Al_{13}^-$ the relevant states are $|A\rangle = |2p, m=\pm 1\rangle$ and $|B\rangle = |1g, m=\pm 1\rangle$. Since their angular momentum quantum numbers differ by $\Delta l = 3$, they cannot interact via $\hat{\tilde{H}}_2'$. (Indeed, the Clemenger-Nilsson diagram[18] shows explicitly that there is no avoided crossing between these two terms when the cluster shape becomes spheroidal.[24])

In cases like this, the transition probability $w$ should be evaluated based on the fact that some additional perturbation must be responsible for mixing the $A$ and $B$ states and facilitating electron transfer into its "home" shell. Interlevel coupling may be supplied, for example, by weak admixtures of other orbital momentum character into the shell wave functions (cf. [6]) and by small-amplitude shape deformations with $L>2$. Let us consider the latter scenario, focusing here on octupolar distortions.

Axially symmetric octupolar deformations are described by the $L=3$ term in Eq. (2). The transition probability is therefore calculated as

$$w = 1 - \exp\left(\frac{-2\pi \langle A|\hat{\tilde{H}}_3'|B\rangle \langle B|\hat{\tilde{H}}_3'|A\rangle}{\hbar \left|\frac{\partial}{\partial t}(E_A' - E_B')\right|}\right), \qquad (19)$$

where $\hat{H}'_3$ is the perturbation Hamiltonian in Eq. (3) with $L$=3. Writing the numerator as a product of two separate matrix elements reflects the fact that perturbation operators $\hat{H}'_L$ are defined in the "squeezed" coordinate system and are thus non-Hermitian. Both terms in $\hat{H}'_3$ contribute to the off-diagonal matrix elements.

The denominator of Eq. (19) makes use of the fact that $vF=\partial E'_{nlm}[\alpha_2(t)]/\partial t$, with the term energies calculated in Eqs. (4),(6). The time dependence $\alpha_2(t)$ is given by Eq. (14), and the derivative is to be evaluated at the time $\tau$ corresponding to the diabatic term crossing point $\alpha_{2,cr}$, Eq. (18).

Since $\hat{H}'_3 \propto \alpha_3$, the octupole deformation amplitude, the transition probability for single passage across the crossing point will be given by

$$w = 1 - \exp\left(-K\alpha_3^2\right), \tag{20}$$

where the factor $K$ contains all the cluster-specific matrix elements and factors in Eq. (19). Its magnitude can be quite large (e.g., for the $Al_{13}^-$ example, it evaluates to $K \approx 2 \times 10^4$), which can make $w$ substantial even for small $\alpha_3$ amplitudes, as shown below.

The octupole shape deformations may be static, or caused by thermal oscillations. To the best of our knowledge, static axial shapes of this type have been considered only for alkali clusters,[25-27] and a few of these have been predicted to have minimal energies for finite, and sometimes even sizeable, values of $\alpha_3$. Such cases imply 100% transition probabilities at the crossing point. However, the calculations are parameter-dependent, and their generality and applicability to other materials has not been accessed.

On the other hand, thermal surface oscillations will always be present in warm clusters in a molecular beam. It makes sense, therefore, to estimate the their contribution to the exponential in Eq. (20). This is described in Appendix B, where the time dependence of $\alpha_3$ and its time-average value are estimated.

Referring again to the case of the $Al_{13}^{-}$ experiment, the value appropriate for use in Eq. (20) is $\alpha_{3,eff}^2 \approx 2 \times 10^{-4}$. Combining it with the aforementioned estimate for $K$, we once again obtain a transition probability close to unity. Consequently, it is reliable to conclude that electron-hole recombination will occur within one or two level crossings, i.e., within a time range of between $\sim\tau$ and $\sim(2\pi\Omega^{-1}\text{-}\tau)$, as calculated at the end of Sec. III. This translates into a range of $\sim$100-200 fs. In other words, within this time interval, the electron will transfer to the lower-shell orbital which it originally vacated as a result of absorbing a photon. Given the approximate nature of the calculation, the result is, in fact, quite consistent with the experimental[3,4] observation of a relaxation time of $\sim$250 fs in the "magic" closed-shell $Al_{13}^{-}$.

As a result of the electronic transition, the cluster now finds itself in the ground electronic state, but with a shape deformed away from the equilibrium. As emphasized earlier, it is essential that this situation represents not a high-order electron-phonon scattering process, but the excitation of a coherent phonon state. The collective distortion will then rapidly dephase into a superposition of incoherent vibrational quanta (i.e., heat). This is an interesting dynamical problem in its own right, but it falls outside the scope of the present paper, since we have seen that the process of electronic relaxation may be considered complete at the level crossing point.

## V. CONCLUSIONS

Time-resolved spectroscopy on free metal clusters has presented a challenge: how is it possible for an excited electron to exhibit very fast relaxation across a shell energy gap which significantly exceeds the vibrational frequencies of the particle? How is the "phonon bottleneck" effect, familiar in nanostructure physics, bypassed in this situation? We have demonstrated that there exists a specific fast electronic relaxation mechanism which involves not a slow multiphonon process, but a fast coherent vibrational excitation: shape deformation of the cluster core. The availability of such a degree of freedom represents a special and distinguishing property of free nanoclusters.

As an application of the theory, the case of the closed-shell $Al_{13}^{-}$ cluster has been considered. The calculated transition time scale provides an explanation for the recent spectroscopic observation[3,4] of surprisingly fast electron-hole recombination in this cluster.

It should be pointed out that the mechanism and formalism discussed here are valid for open-shell (non-spherical) clusters as well. Furthermore, they are applicable to other electronic excitation states and channels involving free clusters: an electron can be injected into an excited energy level in a controlled manner not only by photoexcitation, but, for example, by resonant collisional transfer[28-30] or by the capture of a slow electron. It would be interesting to investigate the relaxation dynamics of such electrons under energy- and time-resolved conditions.

## ACKNOWLEDGMENTS

We are grateful to Dr. V. Z. Kresin for extensive help and discussions. We also thank N. Shevyakina for assistance. This work was supported by a NATO Collaborative Linkage Grant and by the U.S. National Science Foundation under grant No. PHY-0354834.

## APPENDIX A:  THE DISPLACEMENT VECTOR

For shape deformations described by Eq. (2), the velocity of any point in the particle can be derived from a "velocity potential" $\psi$ as $\vec{v} = \nabla \psi$, where[13]

$$\psi = \sum_L \beta_L r^L P_L \left( \cos \theta \right) \tag{A1}$$

and

$$\beta_L = L^{-1} R^{2-L} \dot{\alpha}_L . \tag{A2}$$

The velocity is the time derivative of the displacement vector[19] $\vec{u}$, and therefore we have

$$\vec{u} = \sum_L L^{-1} R^{2-L} \alpha_L \nabla \left[ r^L P_L \left( \cos \theta \right) \right] . \tag{A3}$$

The strain tensor is expressed via Cartesian partial derivatives of $\vec{u}$.[19]  For $L$=2,3 this leads to the results in Section III and Appendix B.

## APPENDIX B:  AMPLITUDE OF OCTUPOLE OSCILLATIONS

The amplitude of $L$=3 shape deformations can be evaluated in a manner analogous to that for quadrupolar oscillations in Sec. III.  For the kinetic and potential energies, one finds (see Ref. [13] and Appendix A, respectively):

$$E_{kin} = \left( 2\pi / 21 \right) \rho R^5 \dot{\alpha}_3^2 , \tag{B.1}$$

$$E_{pot} = \left( 4\pi / 5 \right) R^3 \alpha_3^2 \left( C_{11} - C_{12} + \tfrac{4}{3} C_{44} \right) , \tag{B.2}$$

These are assumed to be small-amplitude thermal oscillations, so the total energy is $E_{kin}$+$E_{pot}$=$E_{thermal}$≅$k_B T$.  (For small oscillations, we can neglect the shift of the electron shell energy.)  The solution of this equation of motion is

$$\alpha_3(t) = b \sin \omega t , \qquad (B.3)$$

with frequency

$$\omega^2 = \frac{42}{5} \frac{C_{11} - C_{12} + \frac{4}{3} C_{44}}{\rho R^2} \qquad (B.4)$$

and amplitude

$$b^2 = \frac{5 E_{thermal}}{4 \pi R^3 (C_{11} - C_{12} + \frac{4}{3} C_{44})} . \qquad (B.5)$$

The effective magnitude of the deformation for use in Eq. (20) can be taken as the time-average of $\alpha_3(t)$, i.e., $\alpha_{3,eff}^2 \approx \frac{1}{2} b^2$.

For $Al_{13}^-$, these relations result in $\omega \approx 4 \times 10^{13}$ s$^{-1}$ and $\alpha_{3,eff}^2 \approx E_{thermal} / (150 \text{ eV})$. Clusters in the experiment in Refs. [3,4] were estimated to be at $T \approx 300$ K, which gives $\tilde{\alpha}_{3,eff}^2 \approx 2 \times 10^{-4}$. This, as anticipated, is a small shape distortion (about 1%), but it gives a serious contribution to the relaxation probability in Eq. (20).

**FIGURE CAPTION**

**Fig. 1.** Scheme of the electronic relaxation mechanism in a free cluster proceeding via a coherent spheroidal shape deformation. The drawing illustrates the example of an electron-hole excitation created in an $Al_{13}^{-}$ cluster.

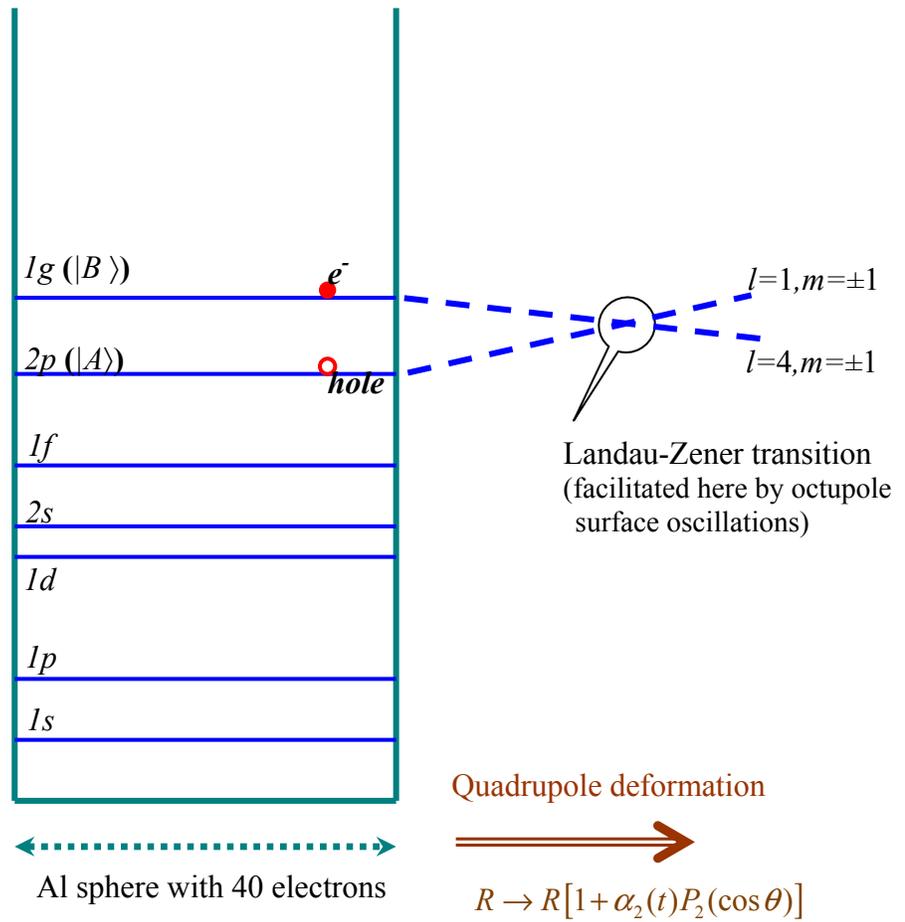

**Fig. 1**